# A signaling protocol for service function localization

Mauro Femminella, Gianluca Reali, and Dario Valocchi

***Abstract*—Current proposals for combining service functions (SFs) do not address some critical management issues, such as the discovery of SF instances close to IP data paths. This information is crucial for deploying complex services both in large cloud networks, where SFs may be moved or replicated, and in the emerging fog/mobile edge computing systems. For this purpose, in this letter we propose the distributed off-path signaling protocol (OSP). We show the protocol functions and demonstrate its scalability and effectiveness by experimental results.**

## I. Introduction and Background

The service function (SF) management has recently been object of research, since it is strictly related to deploying complex services through the so-called network function virtualization (NFV) [1], by combining cloud computing and software defined networks (SDN).

Many SFs can be virtualized, such as security functions, shaping, caching, and middlebox management. In this regard, the IETF Service Function Chaining (SFC) working group has defined an architecture based on SF chaining [6] to provide users with services by using virtualization and SDN [4].

All these activities include the management of network resources distributed over geographical networks, accessible in a virtualized environment [1]. However, SFs discovery, localization, and status retrieval are critical aspects not sufficiently considered in the technical literature, although it is explicitly mentioned in the ETSI specification [1].

Virtualized SF instances in data centers could not be on the IP path of routers connecting two arbitrary communicating entities. Thus, in order to properly chain NFV instances, the use of suitable SFs close to IP data paths is essential for avoiding inefficient data redirections. Fig. 1 shows an example of SF chaining. The blue arrow (Fig. 1.a) indicates the IP path. The localization of the available SFs allows selecting the suitable SF instances to identify the service path (green arrow, Fig. 1.c), which, in turn, maps on the IP path followed by flows belonging to the SF chain (red arrow, Fig. 1.b).

In small settings, a logically centralized orchestrator, knowing the *whole* network topology, could manage the localization and deployment functions. In large networks this approach in not scalable. In this case, a hierarchy of orchestrators could be used, but each of them would serve a portion of the network, and could take sub-optimal decisions. More importantly, multiple services could not be effectively managed. In fact, for this purpose controllers have to be constantly updated about the status of each service instance (e.g. each time a content is cached or evicted from a cache), and signaling congestion could occur. These effects would be even exacerbated both in the novel context of fog/edge computing, which requires distributed management of mini-clouds at the network edge through NFV [8], and in mobile edge computing (MEC), a further emerging technology for 5G networks, which includes base station softwarization. The use of a localized and distributed management protocol for MEC/NFV would allow keeping SFs localized in edge clouds, with a small signaling load and service latency.

M. Femminella and G. Reali are with Department of Engineering, University of Perugia, Italy. D. Valocchi is with University College London, UK. This work is co-funded by EU under the project ARES, supported by GÉANT/GN3plus in the framework of the first GÉANT open call.

In order to identify the suitable chain components and collect their status, we propose the off-path signaling protocol (OSP). It allows localizing resources close to IP data paths. OSP makes use of two main functions: on-path packet interception, used for data path identification, and off-path signaling [2], for SF localization with respect to a data path.

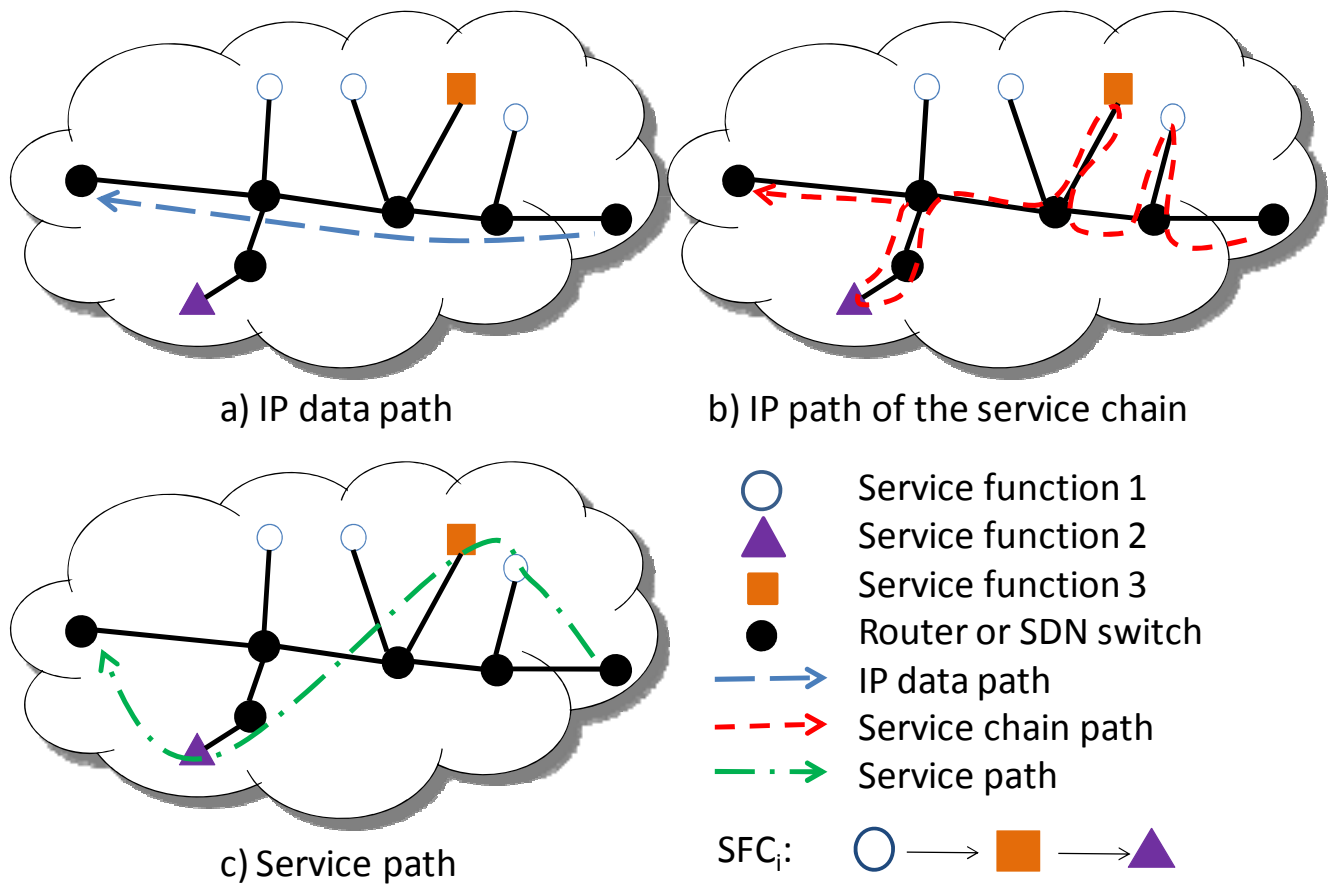

*Fig. 1 – SF chaining in large data networks: a) IP data path, b) IP service chain path, and c) logical service path.*

The existing protocols do not include both these features. For instance, the REsource LOcation And Discovery (RELOAD, IETF RFC 6940) protocol extends the Session Initiation Protocol by introducing peer-to-peer (P2P) features for off-path message exchange. Nevertheless, data path identification is not available. The Next Steps in Signaling (NSIS, IETF RFC 4080) protocol allows storing the state information on the NSIS peers lying on data paths, by leveraging its on-path packet interception capabilities, without supporting off-path signaling. Although an off-path patch was proposed in [3], it was not deployed since it inherits the complexity of the NSIS architecture.

OSP is illustrated in section II, including its state machine diagrams. Experimental results are shown in section III. We draw our conclusion in section IV.

## II. Off-Path Signaling Protocol

The OSP architecture inherits some features of two existing solutions and avoids their shortcomings: NSIS and P2P gossip message exchange [5]. Although it inherits the on-path packet interception from NSIS for implementing off-path operations, its internal state management is highly simplified. Off-path signaling capabilities are protocol native functions. For their implementation, randomized gossiping mechanism for peer discovery are used, as in P2P solutions [5]. The combination of path-coupled operations with off-path signaling allows identifying peers close to data paths. This function is not available in other P2P solutions. Our approach consists of first identifying nodes on the data-path, and then flooding signaling messages from each of them, in order to discover off-path nodes within a maximum distance from the data path. For this purpose, on-path interception is needed. It can be implemented easily, e.g. by port-based filtering or IP options, in SDN devices, software routers, or hardware routers with SDK, thus in any carrier-grade networking platforms.

OSP is organized in two layers. The upper layer, the Signaling Application (SA), implements the signaling logic, and provides a simple interface to the NFV management application. The lower one, the Signaling Transport (ST), distributes SA messages to the intended recipients. In what follows, we describe how these functions are implemented.

### *A. The peer discovery in the Signaling Transport layer*

Peer discovery is a preliminary function used to fill peer tables (PeTs), which are ST data structures used in the off-path signaling distribution described in Section II.B. They include the identity of neighboring peers and the measured IP distance and latency. Each ST node stores in its Peer Table (PeT) the unique peer identifier (PID), the peer IP address, its IP hop distance, the estimated round-trip time, a timestamp of the last gossip session, and a flag indicating its reachability.

OSP is asynchronous and round based. When an ST node is turned on, it only stores the identity of a default node (tracker), used to obtain an initial set of peers. An ST node periodically gossips with the known peers for obtaining further ST node identities, acting as Gossip initiator, by sending Gossip-Registrations. To limit the network overhead, the reachable range of the gossip exchange is set to 1 ST hops. Gossip-Registration messages are intercepted and dropped by the first ST node on the path (the Gossip responder) toward the original destination. It replies back with a Gossip-Response, which is followed by a final Gossip-Ack. Registrations and responses include the identities of some other peers (the *peer to share* list, PTS), randomly selected from the PeT of the initiator/responder, as it typically happens in gossip protocols, such as the Newscast protocol [5].

The protocol operation is illustrated in Fig. 2, which shows also the evolution of PeTs. The initiators are N1 (two times), and then N3. For example, when the first initiator N1 receives the Gossip-Response, it stores the identity of the responder N3 in its PeT, along with the relevant measured metrics. The flag in its PeT entry is set to 1, which indicates that the responder has been *contacted*. In case an ST node (responder) intercepts the Gossip-Registration, the flag associated with the original destination (Tracker in Fig. 2) is set to 1 and its relevant metric values are set to a non-significant value (e.g. -1), i.e. it means that the destination is *out of scope*. The Gossip-Ack notifies the responder that the initiator has received its PTS. By using the information in OSP and IP headers, it is very easy to evaluate the IP distance between initiator and responder. Each peer identity received for the first time in the PTS, or an intercepted, previously unknown, Gossip initiator, has the flag temporarily set to 0 (*uncontacted*, i.e. a not valid metric). N1 is initially set as *uncontacted* in the PeT of N3. The selection of the destination of the next gossip cycle is random, with a higher priority for peers whose flag is 0. Each peer in the PeT is associated with a lifetime to cope with the transient nature of virtualized SFs. If a Gossip-Registration gets no answers, the relevant peer is set as out of scope.

Additional details can be found in [9].

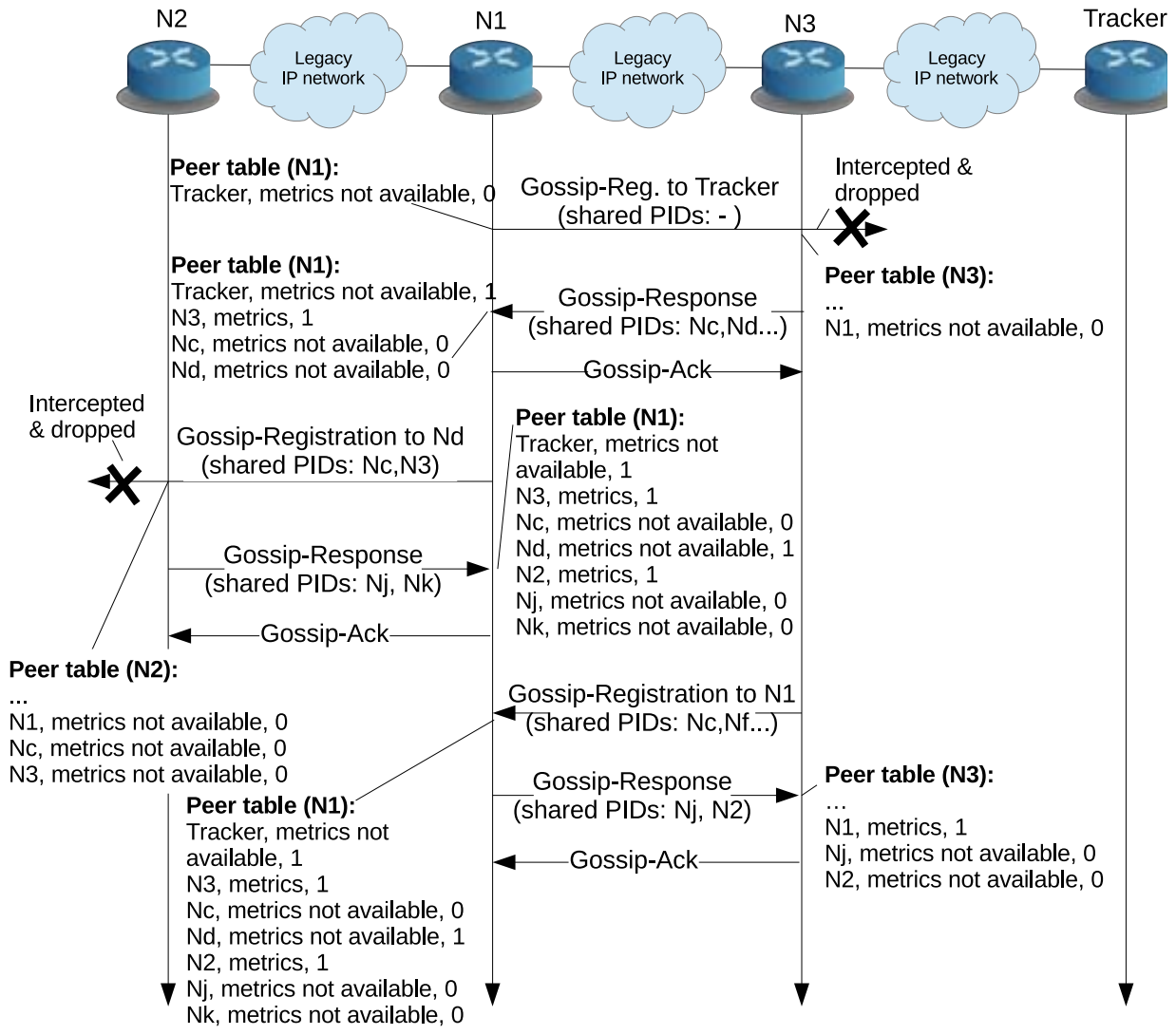

*Fig. 2: Evolution of the gossip-based peer discovery at ST layer.*

### *B. The signaling distribution*

The distribution function is managed jointly by the two layers, ST and SA. Suitable communication primitives confine transport functions at the ST layer, and decision logic at the SA one. Fig. 3 shows their finite state machines (FSMs). In these diagrams, transition edges are labeled with the triggering event (above) and the triggered actions (below). Both SA initiator and forwarder behaviors are modeled by using three states: IDLE, Wait Notification, and Wait Responses. This state definition is flexible enough to both integrate NFV instances easily and introduce multiples SAs protocols. The ST FSM is slightly more complex, since it has to deal with additional lower layer issues, including packet interception and peer selection for signaling distribution.

As for the distribution, we consider an off-path domain which includes the ST nodes staying within a maximum distance *r* (off-path radius) from *at least* one of the nodes of the IP data path. In this paper, we use the IP hop count metric. At the ST layer, the off-path signaling distribution adopts a flooding algorithm, which makes use of two sets of peers:

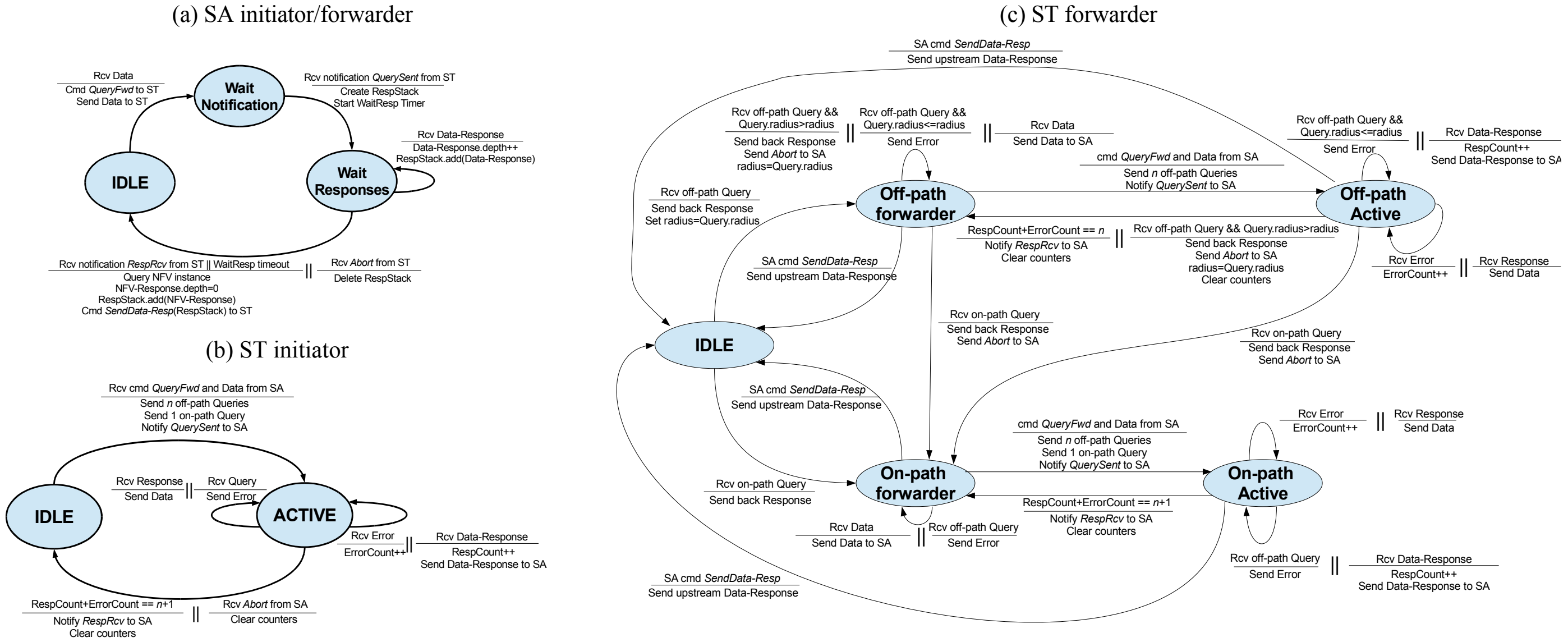

*Fig. 3: State machine for the (a) SA initiator/forwarder,(b) ST initiator and (c) ST forwarder.*

those laying on the IP path between the signaling initiator and destination (on-path peers), and those laying within a distance of radius r IP hops from the path (off-path peers).

*1) Signaling delivery: downstream*

The signaling exchange is initiated by the SA, triggered by an external application (transition from IDLE to Wait Notification in the SA FSM of the initiator, Fig. 3.a). This action triggers a command sent to the ST layer of the initiator (transition from IDLE to ACTIVE in Fig. 3.b). The latter generates an initial query message, by setting both the desired value of the radius *r* and a specific *on-path flag* in the message header of the on-path query, and sends it to the signaling destination. This flag is marked in the signaling messages delivered on the IP data path. Queries are intercepted by the other ST nodes. Each OSP node receiving an on-path query must accept the peering request, send a response message, and be ready to receive SA data from the upstream node (transition from IDLE to On-path Forwarder in Fig. 3.c). When these data are received, they are forwarded to the SA (loop on On-path Forwarder in Fig. 3.c and transition from IDLE to Wait Notification in Fig. 3.a). Then, the SA layer triggers the ST layer to deliver the signaling message not only on-path, but also to the *n* neighbors with a metric value $d \leq r$, namely off-path ST nodes. For each selected off-path peer *i* in PeT with $d_i \leq r$ (Fig. 2), the ST node generates a new query with an updated radius $r\text{-}d_i$ and the on-path flag set to 0 (off-path queries). The new queries are then sent to all the selected peers. The upstream ST node is not selected for off-path distribution. Then, the ST layer notifies the SA and performs a transition towards the Active state (On-path or Off-path, Fig. 3.c), waiting for responses from the queried peers. In turn, the SA FSM moves into Wait Responses (both Fig. 3.a), and creates a stack data structure to store data responses, which are expected within the responses traveling back towards the initiator.

At the ST layer in the Active states, when positive responses are received by the queried peers, SA data are delivered to them (loops on the Active states in Fig. 3.c and on the ACTIVE state in Fig. 3.b).

When a node receives an off-path query, it reads the value of *r*. If $r \geq 1$, the procedure illustrated above is used to select the signaling destinations. The ST state transition is from the IDLE to the Off-path Forwarder (Fig. 3.c). For avoiding the packet duplication problem, typical of flooding algorithms, we have introduced an ST error message: When a forwarder receives a signaling message, it creates an internal soft state for the signaling session, storing the identifier of the served SA protocol, the session identifier, the upstream PID, and the value of *r*. Then, if it receives another ST off-path query from another peer before the time out, with the same set of values and radius $r' \leq r$, it rejects the peering request and sends an error message back to that peer, (error loops on all states except IDLE in Fig. 3.c and Fig. 3.b). In addition, the ErrorCount variable, set to 0 at session setup, is incremented. Instead, if $r' > r$, the previous session is aborted, the SA is notified (transition to IDLE in Fig. 3.a), and a new session is created at ST (transition to Off-path Forwarder, Fig. 3.c).

Similarly, when an off-path node receives an on-path query, it aborts the previous session and establishes a new one, acting as an on-path node. In fact, the value of the radius *r* for an on-path node is always the maximum one, that is that selected by the NFV application that triggers the signaling distribution. In this case, the SA is notified and the relevant states are deleted. The relevant transitions shown in Fig. 3.c are those from off-path states to On-path Forwarder.

*2) Signaling delivery: reverse path*

When a node is a final destination of the off-path signaling and cannot forward the message further, the ST layer moves to the Off-path Active state just to notify the SA layer and then returns into the Off-Path Forwarder state. The SA queries the local NFV instance to get local data, pushes the response into the stack with a depth parameter equal to 0, and triggers the ST to transmit the data response upstream. Session state at SA is cleared, and the FSM returns in IDLE, and the same is done at the ST layer, after sending the data response upstream.

When the ST of an intermediate forwarder receives a data response, it increments the local counter RespCounter, initialized to 0 at session creation, and passes the data response to the SA layer through the relevant APIs (loops on Active states in Fig. 3.b and Fig. 3.c), by also including the metric value *d* of the sending peer, taken from the PeT. The SA pushes this data response into the stack prepared upon entering the Wait Response state, and increases its depth values (loops on Wait Responses state in Fig. 3.a). When all the expected responses are received by the ST layer, that is the number of responses and error messages is equal to *n*, the ST returns back into the Forwarder state in Fig. 3.c, and notifies the SA. The SA queries the local NFV instance, pushes the data into the stack with a depth value equal to 0, and triggers the ST to send the data response stack upstream. Then, it clears all the state variables and moves in IDLE. In turn, the ST sends these data upstream, clears all state information, and moves in IDLE as well. If the WaitResp timer expires at SA, the same actions are executed, without waiting for all the notifications from the ST.

If these events occur at the SA initiator, it sends the collected responses to the querying application going to IDLE.

## III. Performance evaluation

OSP has been analyzed through real experiments. The testbed emulates the Géant network topology, using Linux virtual machines (VM) running in a Gigabit Ethernet cluster. The topology includes 41 routers and 32 servers, used to model points of presence and datacenters, respectively, each implemented by a VM with an OSP instance. We analyzed all the aspects of OSP (peer discovery and signaling distribution).

We begin illustrating the peer discovery analysis, which runs in background (Section II.A). The time needed by each node to discover all its neighboring peers is denoted gossip discovery time ($T_{GD}$). The best value of $T_{GD}$ is achieved with a PTS list of 2 peers, resulting in a mean number of gossip cycles ($n_{GC}$) equal to 36, with $T_{GD}=n_{GC}\times T$, where $T$=5s is the gossip period. This value is about 16 times lower than the $T_{GD}$ of the GIST solution in [3]. The minimum OSP $n_{GC}$ value is the maximum degree of the OSP overlay (10 in the used topology), whereas the maximum OSP $n_{GC}$ is the number of peers of the overlay, except the peer itself ($K$-1=72). As for the overhead, since at each gossip cycle the *i*th OSP node carries out a complete gossip query/response/ack session with one of its neighboring peers, then the *total* average bandwidth overhead is equal to

$$\eta = \frac{1}{T}\sum_{i=1}^{K} v_i \left(G+R+A\right), \qquad (1)$$

where $G$=184 bytes, $R$=184 bytes, and $A$=112 bytes are the size of registration, response, and ack gossip messages, respectively, and $v_i$ is the mean IP distance between the *i*th OSP node and its neighboring peers. With the selected short gossip period (worst case), the OSP gossip discovery produces a negligible signaling bandwidth equal to 55 Kbit/s for the *whole* network (a fraction equal to $3\cdot 10^{-7}$ of the whole network bandwidth), when $v_i$=1, that is OSP is deployed in all network nodes. Additional details can be found in [9].

We now consider the signaling distribution triggered by an NFV application (Section II.B). In all experiments, with all neighbors discovered (worst case), the signaling delivery times are less than 1 s for any path length. Fig. 4 shows the aggregated overhead generated by the signaling distribution as a function of the IP data path length ($L$), for increasing values of $r$. We compare OSP with the GIST proposal [3], since the other gossip algorithms, such as Newscast [5], do not have the requested features (data path proximity control). We measured the overhead at IP layer by using *iptables*, and averaged results over different pairs of peers with the same IP distance. The overhead increases with path length and hose radius. In the worst case ($L$=9, $r$=3), the delivery of a signaling message (message size 1KB) to a very large set of nodes generates only 200 KB of traffic over the whole network. Since a session is completed within 1 second, it requires a fraction of the whole network bandwidth equal to $9\cdot 10^{-6}$, which is negligible. OSP definitely outperforms GIST, with an improvement of about 30-40% for $r$=1, 6 times for $r$=2, and 11-14 times for $r$=3.

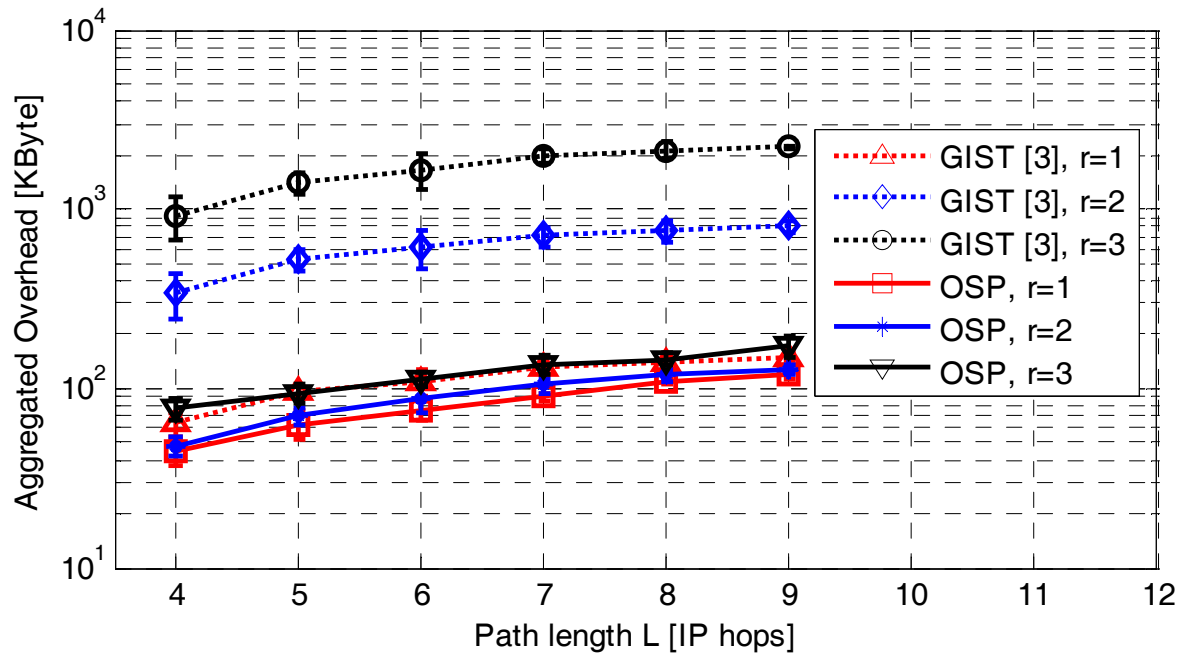

*Fig. 4: Network overhead for NFV probing over an off-path domain of size r (IP hops) vs. IP path length L. Bars indicates 95% confidence intervals.*

## IV. Conclusion

This paper illustrates a new signaling protocol, called OSP, for discovery and localize service functions and make them available for chaining. The original feature of this protocol is its off-path scope, which is enabled through gossip-based discovery and flooding-based distribution. OSP has been implemented and analyzed experimentally. It exhibits the desired features at the expenses of a negligible network overhead. Future work will consider integration with MEC platforms in 5G networks.

# A signaling protocol for service function localization
## Supporting document

Mauro Femminella, Gianluca Reali, and Dario Valocchi

## I. MOTIVATIONS

The underlying framework is that of service functions (SFs) management and their virtualized deployment (network function virtualization, NFV) usable for realizing SF chains to implement complex services. In order to identify the suitable chain components and collect their status, we propose the off-path signaling protocol (OSP). It allows localizing resources close to IP data paths. OSP makes use of two main functions: on-path packet interception, used for data path identification, and off-path signaling, for SF localization with respect to a data path. The main advantage of this solution consists of using a signaling protocol able to take into account not only functional issues (type of NFV instance) but also the network locations where they are instantiated at any given time (i.e. coupled with data plane). It provides the network/service operator with the possibility of better chaining different NFVs for deploying advanced services. This is essential for avoiding inefficient data redirections in service composition, and translates into decreasing the volume of the exchanged traffic.

In order to execute the proposed resource discovery protocol, it is necessary to make use of preliminary peer discovery system, based on gossiping the available peer information due to its intrinsic robustness, with an estimation of the relevant topological information (i.e., peer distance measurement). This is needed to implement the path-decoupled delivery mechanism, which is the basis of our proposal of SFs discovery and management. In other words, an original contribution of our approach with respect to other known proposals, such as the Newscast one [1], is to add topological information to the membership management, used to implement off-path communication capabilities. Actually, Newscast is a very interesting solution, and we admittedly inspired to it for designing some features of our proposal. Our gossip-based peer discovery protocol and Newscast shares some aspects, especially those regarding the basis of the membership management. Their main difference relies in the fact that the simple and efficient membership management system natively provided in Newscast does not take into account the relative distance between members and does not provide metric measurement. In fact, from our understanding of the Newscast model, defining a subset of nodes acting as receivers for a certain information, based on topological parameters, is not a trivial task. The distance measurement, which comes embedded with our peer discovery system, is a fundamental feature, needed to implement the path-decoupled signaling delivery mechanism.

These differences lead to the need of a custom peer discovery and membership management mechanism, which is implemented in the lower layer of OSP. Thus, the requirements of the above signaling layer (off-path signaling distribution) imposes the design of a novel lower layer discovery protocol which runs in background and which is aware of the relevant position of neighboring peers. The signaling distribution uses exactly this information stored in the peer tables (PeTs) during the distribution phase, thus a classic peer sampling algorithm would not be suitable.

## II. OFF-PATH SIGNALING PROTOCOL

### A. *The Peer Discovery Solution*

Peer discovery is a preliminary function used to fill peer tables (PeTs), which are signaling transport (ST) data structures used in the off-path signaling distribution. They include the identity of the neighboring peers and their measured IP distance and latency. Each ST node stores in its Peer Table (PeT) the unique peer identifier (PID), the peer IP address, its IP hop distance, the estimated round-trip time, the timestamp of the last gossip session, and a boolean flag indicating its reachability. An example of a PeT is reported in Table 1.

OSP is asynchronous and round based, with a gossip period of size T. When an ST node is turned on, it only stores the identity of a default node (tracker), used to obtain an initial set of peers. The identity of tracker is a configuration parameter. An ST node periodically gossips with the known peers for obtaining further ST node identities, acting as Gossip initiator, by sending Gossip-Registrations. To limit the network overhead, the reachable range of the gossip exchange is set to 1 ST hops. This means that Gossip-Registration messages are intercepted and dropped by the first ST node on the path (the Gossip responder) toward the original selected destination. The gossip responder replies back to the gossip initiator with a Gossip-Response, which is followed by a final Gossip-Ack. Registrations and responses include the identities of some other peers (the peer to share list, PTS), randomly selected from their respective PeT, as it typically happens in gossip protocols, such as the Newscast protocol [1].

In more detail, the peer table is managed by dividing the peers into three subsets/lists:

- *Neighbor peers*, which lies 1 OSP hop away from the node and which has been involved in at least one gossip session with the node (i.e. the PeT contains the peer metrics, the flag set to 1).
- *Unreachable peers* (out of scope peers), which have been selected as destination for a gossip session, but are at a distance greater than 1 OSP hops (i.e. metrics are set to a non-significant value, such as -1, since they cannot be evaluated, although the flag set to 1).

*Table 1. Example of PeT.*

| # | PeerID | IP Address | Metrics | | Timestamp | Flag |
|---|---|---|---|---|---|---|
| | | | IP Hops | Latency | | |
| 1 | kJNg | 10.0.3.1 | / | / | 1410704001 | 0 |
| 2 | p2uQ | 10.0.32.2 | 2 | 400 | 1410704003 | 1 |
| 3 | AuSp | 10.0.223.1 | -1 | -1 | 1410704005 | 1 |

- Unknown peers (uncontacted peers), the identity of which has been received during a gossip session although they still need to be selected as destinations for a gossip session (i.e. metrics are not set and the flag is set to 0).

The list of *unknown peers* is increased during each gossip session, by appending to the list the shared identities that were not previously included in the PeT. Once a peer has been involved in a complete gossip exchange, its identity is moved in the *neighbors* list or in the *unreachable peers* list, depending on the outcome of the gossip signaling exchange. In more detail, if the gossip responder is the queried node (gossip destination), it is classified as *neighbor*, otherwise, if the gossip responder is a node on the path intercepting the Gossip-Registration, the gossip destination is classified as *unreachable/out of scope*. The destination of the gossip session, i.e. the peer to gossip, is selected randomly among the so-called *unknown peers*, if present. This strategy allows completing the network discovery phase quickly, by classifying as soon as possible all peers as *neighbor* or *unreachable*. Otherwise, if the *unknown peers* list of a given OSP node is empty, that is it has already tried to contact all discovered peer identities and has classified them as *neighbor* or *unreachable*, the selection of the peer to gossip becomes completely random, considering both the *neighbor* or *unreachable peer* list, without any priority. This allows taking into account also topology/routing changes.

Storing *unreachable peers* allows a node to avoid wasting gossip sessions to contact nodes that are known to lie beyond its scope, at least during the discovery phase. The size of the *unreachable peers* list is configurable and can be managed with a standard LRU algorithm. We tested the impact of the size of this list on the convergence time of the peer discovery protocol. As expected, following the decrease of the size of this list, the performance of the protocol (i.e. the discovery time) slightly decreases. In fact, the unreachable peer identities are evicted from the table more frequently and are inserted again as *unknown peers* during gossip sessions, thus slowing the protocol convergence. However, since this performance penalty is negligible for reasonable sizes of *unreachable peers* list, we did not limit the number of *unreachable peers* to store, since their number can, at most, be equal to the number of OSP peers in the considered network. Hence, the memory requirement is affordable.

Also *neighbor peer* entries are maintained as soft statuses, and the lifetime of each identity depends on the size of the neighbor list and on the gossip timer. Although this mechanism is tunable through a configuration parameter, the best approach is to design the OSP protocol to be self-adaptive, so as to increase/decrease the PeT entries lifetime on the basis of the current size of the peer list. In this way, it is easy to ensure a gossip attempt to any stored peer identity before their lifetime expiration.

As a consequence of a three way gossip exchange, in the PeT of the gossip initiator, the gossip responder is flagged as a *neighbor peer*, with valid peer metric (e.g., see peer 2 in Table 1), whereas the gossip destination is flagged as an *unreachable peer* (e.g., see peer 3 in Table 1). In fact, although the gossip initiator tried to contact the gossip destination, it has received a response by the gossip responder, which is a different node. This means that the gossip destination is more than 1 OSP hop away from the gossip initiator, and the gossip responder is the first OSP node on the path connecting gossip initiator with the gossip responder. Instead, in the PeT of the gossip responder, the identity of the gossip initiator is set as an *unknown peer* (e.g., see peer 1 in Table 1), since its identity was obtained during a gossip exchange, but it has still not tried to contact it directly in order to measure network metrics, which are necessary for signaling distribution.

Upon the transmission of a Gossip-Registration, as typically done for managing soft-states, a gossip timer is issued. Hence, the gossip protocol is purely cycle-based, so if a gossip message is not received correctly by the timeout, the session is aborted, the gossip destination statuses are left unaltered in the PeT, the protocol statuses are reset, and the initiator node waits for the subsequent gossip cycle to start a brand new gossip session. Thus, the length of the gossip timer is a parameter which can be tuned in order to adapt to different application requirements. Our choice is to set the gossip timer large enough to ensure the completion of the gossip session exchange before the next gossip round, unless a packet loss occurs. In particular, we set the gossip timer equal to the gossip period, whose length is reasonably equal to a few seconds. In this way, there is plenty of time for involved entities to complete the gossip exchange, also in case of heavy overload.

The formal definition of the proposed protocol is illustrated by the finite state machine of gossip initiator and gossip responder, shown in Fig. 1 and Fig. 2, respectively.

The packet format can be represented in ABNF as illustrated in Fig. 3.

### B. *The signaling distribution*

In this section, we report the ABNF format of messages used to implement signaling distribution at the Signaling Transport (ST) and the Signaling Application (SA) layers, illustrated in Fig. 4 and Fig. 5, respectively.

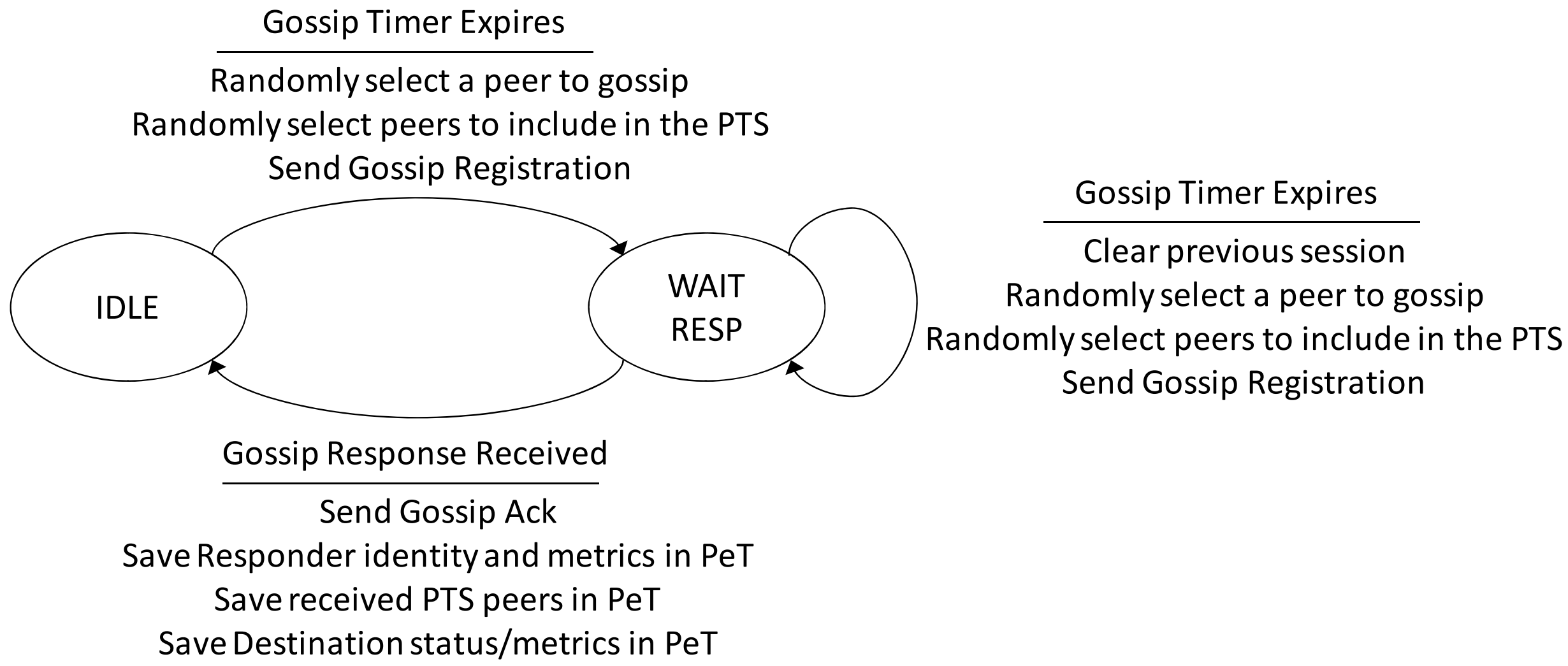

*Fig. 1: FSM for the gossip initiator.*

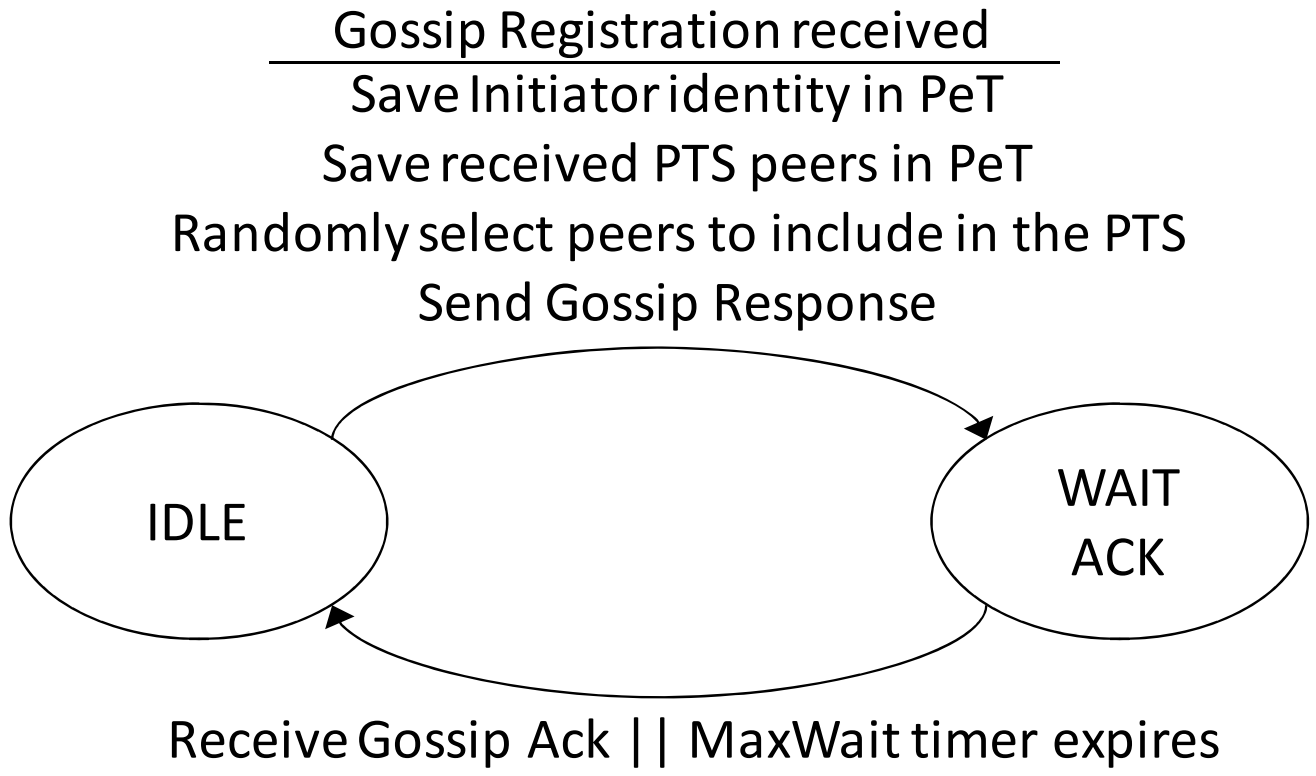

*Fig. 2: FSM for the gossip responder.*

```
Registration    =       Message Type
                        Source Peer-Identity
                        Destination Peer-Identity
                        Source IP address
                        Session-Identifier
                        Metric value
                        (Shared Peer Ids)

RegResponse     =       Message Type
                        Source Peer-Identity
                        Destination Peer-Identity
                        Source IP address
                        Session-Identifier
                        Metric value
                        (Shared Peer Ids)

Ack             =       Message Type
                        Source Peer-Identity
                        Destination Peer-Identity
                        Session-Identifier

Shared Peer Ids =       Peer-Identity
                        IP address type
                        IP address
```

*Fig. 3: ABNF format for gossip packets.*

## III. Implementation issues

### A. Protocol implementation

Our approach consists of first identifying nodes on the data-path, and then flooding signaling messages from each of them, in order to discover off-path nodes within a maximum distance from the data path. For this purpose, on-path interception is needed. Our implementation of packet interception in the lower layer of OSP can be based on different technologies. In any case, it does not require to have the NICs of the various peers configured in promiscuous mode.

For instance, it is possible to use the Router Alert Option (RAO) field of the IP protocol to instruct the on-path agents to process a packet when it is processed by using their forwarding table. In particular, our implementation, which makes use of software routers and servers, combines the IP RAO field and of the Netfilter library for Linux system to implement this mechanism. Each node running the OSP agent has a set of Netfilter rules which append the packets that match a pre-determined UDP port to the protocol daemon queue. When a new node connects to the system, it sends the first registration packet, which is a UDP packet, with a destination IP equal to the IP address of the tracker. The first node on the path between the new node and the tracker makes use of this rules to filter the OSP packets and append them to the OSP queue. Then, the protocol daemon checks the IP header of the packet to verify if the correct RAO flag is set and, in this case, it processes the packet. This is an easily feasible approach, since modern software routers may have high end performance, comparable with those of hardware devices, as shown in [4][5].

In addition, given the popularity and widespread adoption of the SDN paradigm [7], it is possible to envisage another implementation of packet interception. The OSP entities could run on a stub node (i.e. a server),which is connected to an SDN switch. The OSP daemon could thus register with the SDN controller, by using it to set the forwarding rules for the OSP UDP and TCP ports into the switch. Each packet arriving to the switch which matches the OSP rules will be forwarded to the OSP stub interface, which uses the Netfilter rules to append the packet to the daemon queue, in order to process it.

Then, if the packet needs to be forwarded, the processed packet can be sent back to the switch, which uses the standard rule set to forward it to the next hop. Since the forwarding plane is separated by the control plane, essentially the OSP protocol does not have any significant impact on the forwarding performance.

Finally, the OSP protocol can be also easily implemented in all hardware routers that are equipped with a software development kit [6], since its only layer 3 requirement is the packet interception, and all the other functions runs at the application layer.

Thus, we can conclude that the OSP protocol can be implemented easily in SDN devices, software routers, or hardware routers with software development kit, thus in any carrier-grade networking platforms. On all these platforms, the impact of the OSP protocol on the forwarding performance, given its negligible requirement in bandwidth consumption, is really negligible.

### B. Transport layer issues

The OSP Messages can be transported by any layer 4 protocol. This section illustrates sensitivity to packet losses and overhead of gossip-based peer discovery and signaling distribution in case either UDP or TCP is used.

#### 1) Gossip-based peer discovery

Peer discovery is a process based on gossiping between peers. Since it consists of three messages (Registration, RegResponse, and Ack, as shown in Figure 3), the most reasonable choice is to use UDP, due to the high overhead in establishing a TCP session for exchanging just three small application layer messages. The gossip mechanism, which adopts soft states, cancelled when their associated timer expire,, is intrinsically robust to packet losses. Any session experiencing a packet loss, thus unable to complete the three-way handshake, will be attempted again in a subsequent gossip cycle.

#### 2) Signaling distribution

The signaling distribution is a more complex process, and can use both pure UDP and mixed UDP/TCP. In any case, the OSP protocol uses timer-based soft states also in the signaling distribution. This means that that a session cannot be locked by a packet loss, but, at most, some part of the overlay distribution tree cannot be covered.

If UDP is used, it could happen that one of the messages listed in Fig. 4 is lost. However, this does not necessarily means that a portion of the network cannot be reached by signaling messages. In fact, due to the flooding-based operation of the OSP protocol, most of the potential losses can be compensated by additional Queries, arriving from a different path. If the Response is lost, the queried node could send an Error message upon receiving a new Query, based on the comparison between the stored “radius” value and the “radius” value present in the Query message. Finally, if Data or Data-Response messages are lost, the initiator cannot distribute or receive information from a portion of the overlay distribution tree.

Also the TCP protocol can be used to transport signaling distribution messages. However, due to the flooding-based operation of the signaling distribution, a significant number of Query + Error exchanges could happen. Since the overhead for managing a TCP session to just exchange a Query+Error messages is excessive, it seems more reasonable to set up the TCP session only between those peers that have established a “peering agreement” (Query followed by a Response, both transported over UDP). In this case, TCP guarantees the reliable delivery of larger Data and (stacked) Data-Response messages. Also, the number of TCP sessions would be much lower, with a significant overhead reduction. In addition, in this way it is ensured the Data-Responses coming from the leaves of the distribution tree is reliably delivered when they receive Data messages.

```
Query                   =     Message Type
                              Source Peer-Identity
                              Destination Peer-Identity
                              Source IP address
                              Destination IP address
                              Session-Identifier
                              On-path flag
                              Metric Type
                              Radius
                              SA-Identifier
Response                =     Message Type
                              Source Peer-Identity
                              Destination Peer-Identity
                              Source IP address
                              Destination IP address
                              Session-Identifier
                              SA-Identifier
Error                   =     Message Type
                              Source Peer-Identity
                              Destination Peer-Identity
                              Session-Identifier
                              Source IP address
                              Destination IP address
                              Error code
Data                    =     Message Type
                              Source Peer-Identity
                              Destination Peer-Identity
                              Source IP address
                              Destination IP address
                              Session-Identifier
                              SA-Identifier
                              SA-Payload
Data-Response           =     Message Type
                              Source Peer-Identity
                              Destination Peer-Identity
                              Source IP address
                              Destination IP address
                              Session-Identifier
                              SA-Identifier
                              SA-Payload
```

*Fig. 4: ABNF format for signaling delivery packets at ST layer.*

```
Setup                   =     Message Type
                              Service Type
                              [SF Payload]
Remove                  =     Message Type
                              Service Type
                              [SF Payload]
Probe                   =     Message Type
                              Service Type
                              [SF Payload]
Response                =     Message Type
                              Response code
                              *(SF Status Element)
SF Status Element       =     Node-Identifier
                              Status code
                              Depth
```

*Fig. 5: ABNF format for signaling delivery packets at SA layer.*

Finally, as it happens in the UDP case, the flooding-based operation guarantees the coverage of most of peers through multiple paths, thus limiting the impact of packet losses in Query messages.

## IV. Performance Evaluation

OSP has been analyzed through real experiments. The testbed emulates the Géant network topology, by using Linux virtual machines (VM) running in a Gigabit Ethernet cluster. The topology modeled both 41 points of presence in the Géant network, by using 41 software routers, and 32 data-centers, by using 32 Linux servers. Thus, not only it is representative of a real network, but also takes into account the server-based implementation of NFV in data-centers. Both servers and software routers have been implemented by an VM with a running OSP instance.

We begin illustrating the discovery time for the peer discovery analysis, which runs in background. The time needed by each node to discover all its neighboring peers is denoted gossip discovery time ($T_{GD}$). The best value of $T_{GD}$ is achieved with a PTS list of 2 peers, resulting in a mean number of gossip cycles ($n_{GC}$) equal to 36, with

$$T_{GD}=n_{GC}\times T, \quad (1)$$

where $T$=5s is the gossip period. This choice of the gossip period allows completing all gossip exchanges by the timeout, and, at the same time, limiting the time needed to complete the discovery. Any increase or decrease of the gossip period, as shown by (1), just implies a linear effect in the gossip discovery time. Only if the $T$ value becomes too small (few ms), it can happen that a timeout occurs when a session is still in progress. However, a value of a few seconds is recommended, also to limit the network overhead, as shown in what follows.

The $T_{GD}$ value is about 16 times lower than the $T_{GD}$ of the GIST solution in [3]. The minimum OSP $n_{GC}$ value is the maximum degree of the OSP overlay (10 in the used topology), whereas the maximum OSP $n_{GC}$ is the number of peers of the overlay, except the peer itself ($K$-1=72). The impact of the size of the PTS on the $T_{GD}$ is shown in Fig. 6. We have performed multiple runs for each value of the PTS, due to the random selection of PID shared in the PTS list. The confidence intervals are indicated in the figure as well. As we can see, initially the $T_{GD}$ decreases by increasing the PTS from 1 to 2 peers. This is expected, since if a larger number of peer identities is provided, it is more likely that no gossip cycles in which the *unknown peer* list is empty exist, but the discovery of all neighbors is still incomplete. However, the increase of the number of PID in the PTS, not only does not imply any improvement, but causes a performance degradation. This happens since sharing many peer identities enlarges the set of peers selectable as gossip destination for the next gossip session. In addition, the number of possible gossip destinations is much higher than the number of next hop OSP routers, which are the neighbors to discover. When these routers are OSP nodes, it is disadvantageous to test a large set of peers, most of which are unreachable due to the 1-hop scope limitation. When the paths to most of peers share the same next hop, the interception of other next hop routers happens less frequently, thus enlarging, on average, the discovery times. This leads to the conclusion that sharing few peer identities is convenient both to limit the signaling overhead of the peer discovery mechanism and to reduce the convergence time.

In any case, the obtained values are always within the above maximum and minimum values. These bounds are due to the specific policy used by the OSP protocol to select the peer to gossip. In fact, when selecting the peer to gossip, the protocol gives priority to *unknown peers*, thus trying to contact all peers before sampling again *neighbors* or *unreachable* ones. Instead, the completely random selection policy adopted in the GIST-based solution in [3] implies much worse performance, which are nearly independent of the number of PID shared in the PTS. In order to investigate if this phenomenon is either due to the selected network topology or it is more general, we have repeated the same experiments in other two well-known real network topologies: the Abilene topology, consisting of 34 nodes [2], and the Deltacom topology, consisting of 113 nodes [2]. Results are shown in Fig. 7 and Fig. 8, for Abilene and Deltacom topologies, respectively. Basically, the same considerations done for Fig. 6 hold. Finally, it is worth noting that the best, average performance (i.e. that obtained with a number of PID shared in the PTS list equal to 2) is always closer to the lower bound than to the upper bound.

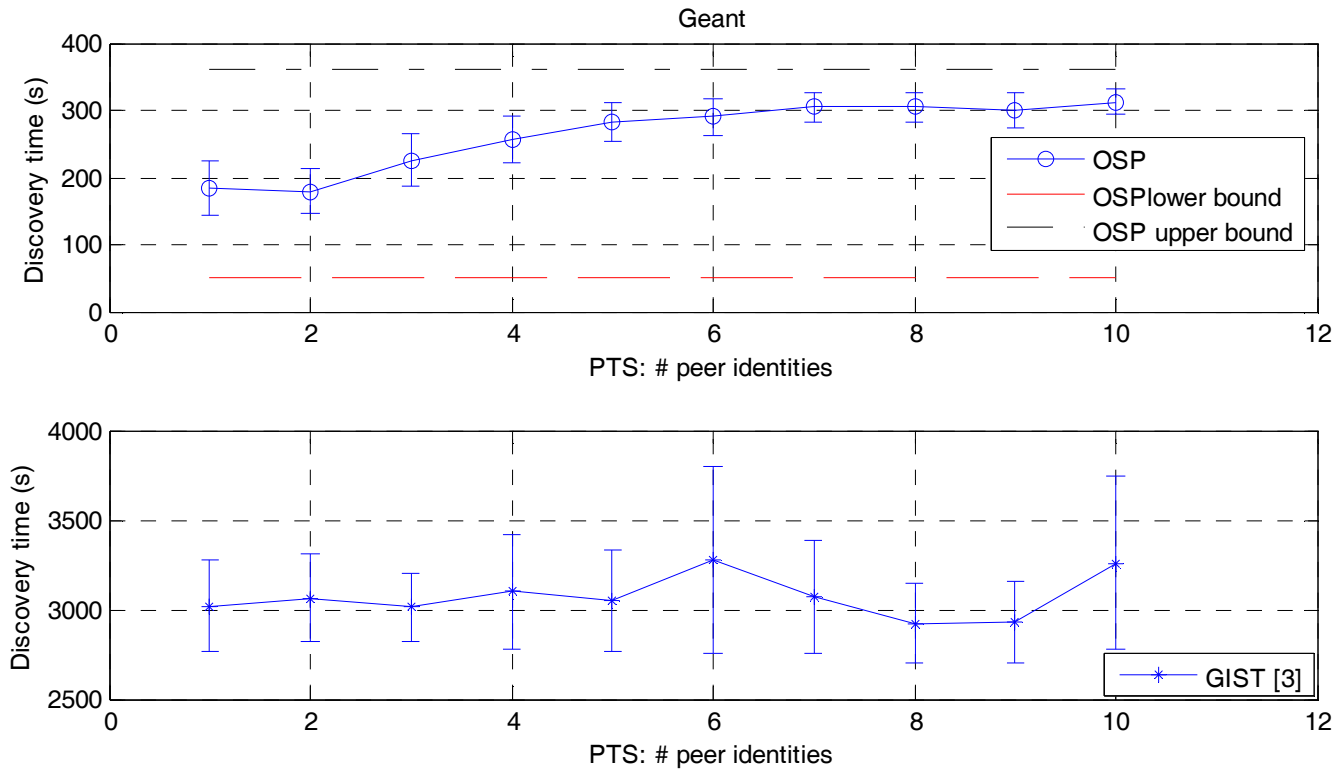

*Fig. 6: Discovery times vs. the number of shared peer identities (size of PTS list) for the OSP gossip algorithm and the GIST-based solution in [3]. Géant topology [2].*

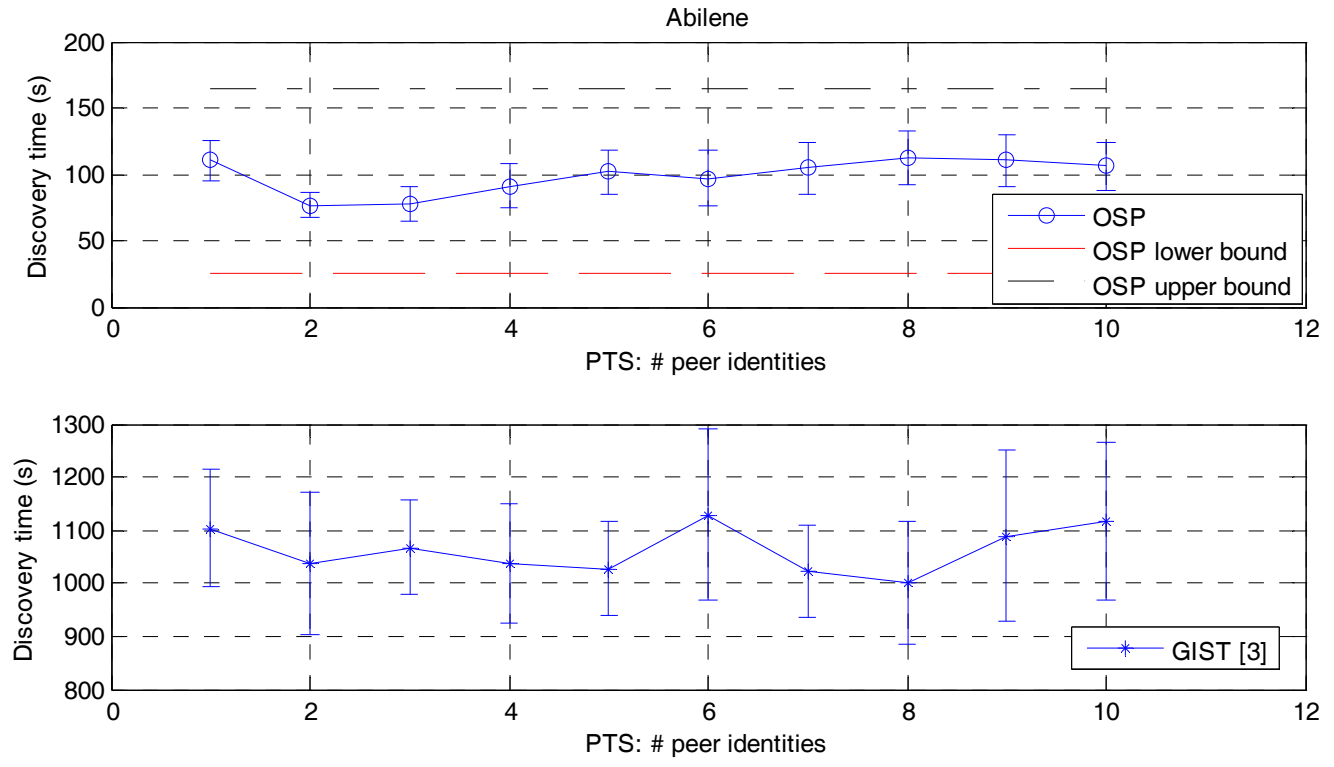

*Fig. 7: Discovery times vs. the number of shared peer identities (size of PTS list) for the OSP gossip algorithm and the GIST-based solution in [3]. Abilene topology [2].*

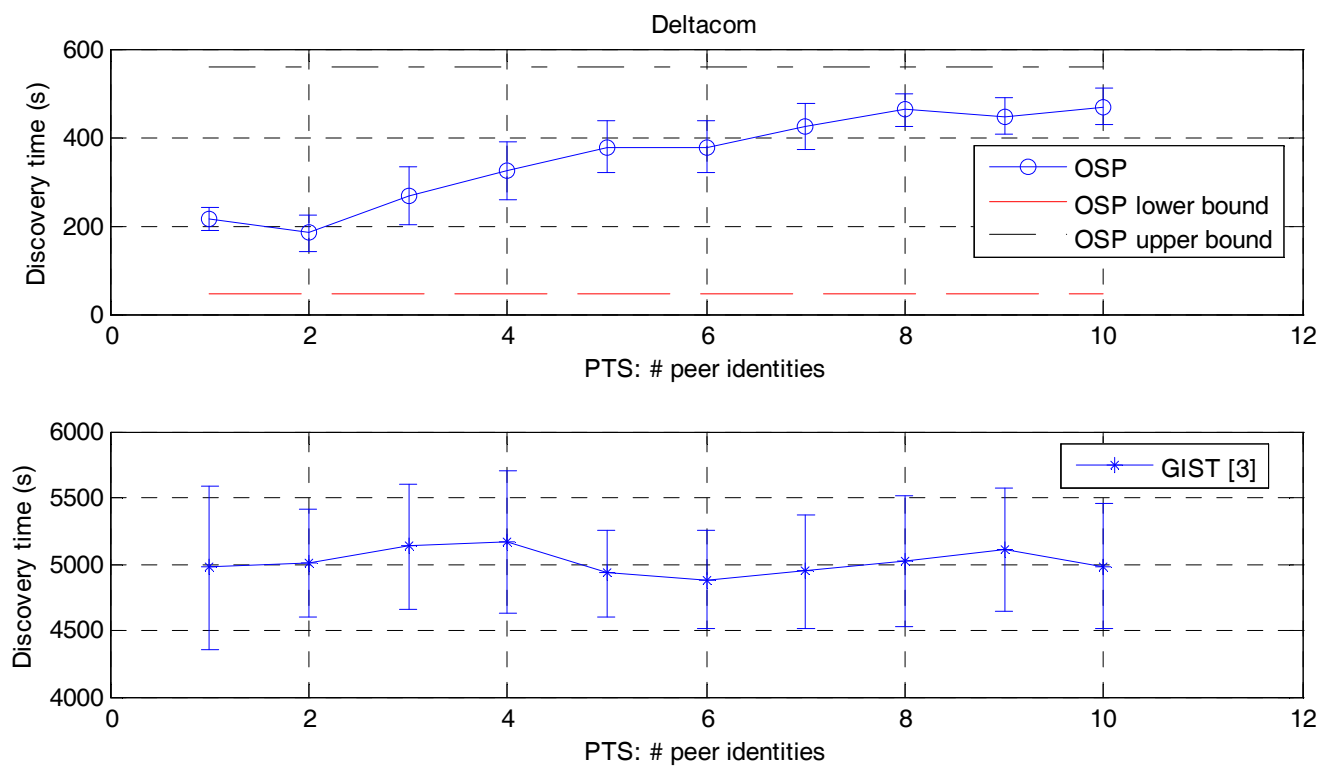

*Fig. 8: Discovery times vs. the number of shared peer identities (size of PTS list) for the OSP gossip algorithm and the GIST-based solution in [3]. Deltacom topology [2].*

As for the overhead, since at each gossip cycle the *i*th OSP node carries out a complete gossip query/response/ack session with one of its *neighboring peers*, then the total average bandwidth overhead is equal to

$$\eta = \frac{1}{T}\sum_{i=1}^{K} \nu_i (G + R + A), \qquad (2)$$

where $G$=184 bytes, $R$=184 bytes, and $A$=184 bytes are the size of Registration, RegResponse, and Ack gossip messages, respectively, and $\nu_i$ is the mean IP distance between the ith OSP node and its neighboring peers. With the selected short gossip period $T$ (worst case from a bandwidth consumption viewpoint), the OSP gossip discovery produces a negligible signaling bandwidth equal to 55 Kbit/s for the whole network with the Géant topology (a fraction equal to $3 \cdot 10^{-7}$ of the whole network bandwidth), when $\nu_i$=1, that is OSP is deployed in all network nodes. The bandwidth consumption becomes about 27 Kbit/s for the whole network with the Abilene topology, and 85 Kbit/s for the whole network with the Deltacom topology.

These numbers are relevant to 2 PID carried in the PTS. Clearly, by increasing the size of the PTS, the values of $G$ and $R$ increase as well. However, since there is no reason to use larger values, these are representative of the typical bandwidth consumption.

Finally, it is worth considering an incremental OSP adoption in the network. From (2), two contrasting contributions arise. Decreasing the number of OSP peers means both decreasing $K$ and increasing $\nu_i$. We tested the introduction of partial deployment of OSP peers on the Géant topology, by considering also 25%, 50%, and 75% of nodes (servers/routers) running OSP, and averaging test results obtained with different configurations. Results are reported in Fig. 9. First of all, it is important to note that that the overall effect is low. In any case, by decreasing the number of OSP nodes from 100% to 75% implies a small increase in overhead ($K$ is still high, but also $\nu_i$ increases). When $K$ is further decreased, the increase in the average OSP peer distance ($\nu_i$) is not able to compensate it, and overhead significantly decreases. Thus, a partial introduction of OSP does not impact on system performance.

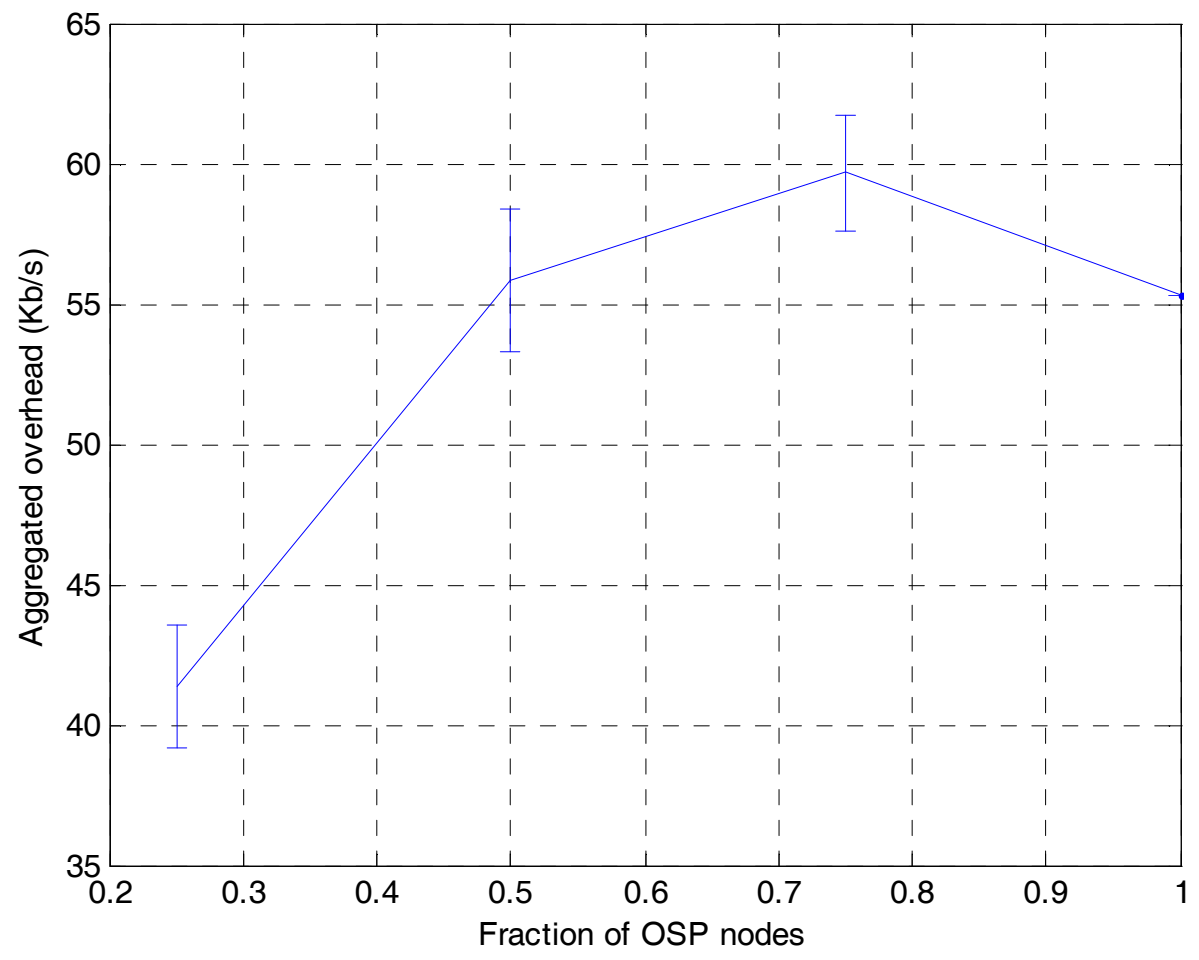

*Fig. 9: Aggregated overhead of gossip discovery vs. fraction of OSP nodes. Géant topology [2].*